\documentclass[12pt]{article}
\usepackage{epsf}

\newcommand{\eeq}{\end{equation}}

\newcommand{\beq}{\begin{equation}}
\newcommand{\beql}{\begin{eqnarray}}
\newcommand{\eeql}{\end{eqnarray}}

\begin{document}

\title{ Glassy Random Matrix Models }

\author{N. Deo\\
Poornaprajna Institute of Scientific Research,\\
Sadashiva Nagar, Bangalore 560080, India}

\maketitle

\begin{abstract}

This paper discusses Random Matrix Models which exhibit the 
unusual phenomena of having multiple solutions at the
same point in phase space. These matrix models have gaps in 
their spectrum or density of eigenvalues. The free energy and 
certain correlation functions of these models show differences 
for the different solutions. Here I present evidence for 
the presence of multiple solutions both analytically and numerically.

As an example I discuss the double well matrix model 
with potential $V(M)= -{\mu\over 2} M^2 + {g\over 4} M^4$ where 
$M$ is a random $N\times N$ matrix (the $M^4$ matrix model)
as well as the Gaussian Penner model with 
$V(M) = {\mu\over 2} M^2 - t \ln M$. 
First I study what these multiple solutions are in the large N limit 
using the recurrence coefficient of the orthogonal polynomials.
Second I discuss these 
solutions at the non-perturbative level to bring out some 
differences between the multiple solutions.  
I also present the two-point density-density correlation functions which
further characterizes these models in a new universality class. A motivation 
for this work is that variants of these models have been conjectured to 
be models of certain structural glasses in the high temperature phase. 
 
\end{abstract}

PACS: 02.70.Ns, 61.20.Lc, 61.43.Fs


\section{Introduction}\label{intro}

Matrix models are known to be of 
importance in many diverse areas, for example,
in quantum chaos, disordered condensed matter systems, two-dimensional
quantum gravity, quantum chromodynamics and strings. In the context of 
the one-matrix models, the models that have been studied correspond
to an eigenvalue distribution on a single-cut in the complex plane where the 
eigenvalue density is non-zero \cite{mg}. An obvious generalization is
to study a matrix model with a more complicated 
eigenvalue structure. Here I study a class of models with a
single hermitian matrix
but with two cuts for the eigenvalue density and point out some
unusual features of these models different from the single-cut
matrix models. One of the important differences observed in these 
models is that they have multiple solutions whose effect appears in
certain correlation functions. This work suggests
the possibility that these multiple solutions may be
glassy and hence these models may be useful in describing certain real 
disordered glassy systems ref. \cite{ckpr,p}.  

I study here the matrix model with a double-well potential as well 
as the Gaussian Penner model where similar results are obtained
ref. \cite{bdjt}. In both cases the
potential is symmetric about the origin. In addition to the usual
symmetric solutions I discuss the symmetry breaking solutions.
I study these solutions in the large $N$ limit and then 
at the non-perturbative level. I also calculate 
the two-point density-density correlators using
orthogonal polynomial methods and methods of steepest descent
which strengthens the above observations and classifies
the models in a new universality class \cite{ajm,bz}. 

\section{Notations and Conventions}\label{note}

Let $M$ be a $N\times N$ hermitian matrix. The partition function 
to be considered is $ Z=\int dM e^{-N Tr V(M)} $. The Haar measure 
is given by $dM = \prod_{i=1}^{N}dM_{ii}\prod_{i<j} dM_{ij}^{(1)}
dM_{ij}^{(2)}$ where $M_{ij}=M_{ij}^{(1)} + i M_{ij}^{(2)}$; 
there are $N^2$ independent variables. $V(M)$ is a polynomial in $M$:
$V(M)=g_1 M + (g_2/2) M^2 + (g_3/3) M^3 + (g_4/4) M^4 + \ldots$ . $Tr V(M)$
and the measure $dM$ are invariant under the change of variables
$M\rightarrow M^{\prime}=U M U^{\dagger}$ where $U$ is a unitary matrix. 
We can use this invariance to express $Z$ as an integral over the 
eigenvalues $\lambda_1,\lambda_2,...,\lambda_N$ of $M$.
Writing $M= U D^{\prime} U^{\dagger}$ where $D^{\prime}=diag(\lambda_1,
\lambda_2,.....\lambda_N)$, the partition function becomes
$ Z = \int dU \int_{-\infty}^{\infty} \prod_{i=1}^{N} d\lambda_i 
\Delta (\lambda)^2 e^{-N \sum_{i=1}^N V(\lambda_i)} $
where $\Delta (\lambda) = \prod_{i<j} |\lambda_i-\lambda_j|$
is the Vandermonde determinant arising from the change of variables. 
The integration $dU$ is trivial because the integrand is independent of
$U$ due to the invariance, and gives a constant factor. 
By exponentiating the determinant one arrives at the 
Dyson Gas or Coulomb Gas picture. Thus the partition function is 
$
Z = C \int_{-\infty}^{\infty} \prod_{i=1}^N d\lambda_i e^{-S(\lambda)} 
$
where
$ S(\lambda) = N \sum_{i=1}^N V(\lambda_i) - 2
\sum_{i\ne j} ln |\lambda_i-\lambda_j|$, and $C$ is a constant. 

\begin{figure} 

\leavevmode
\epsfxsize=4in
\epsffile{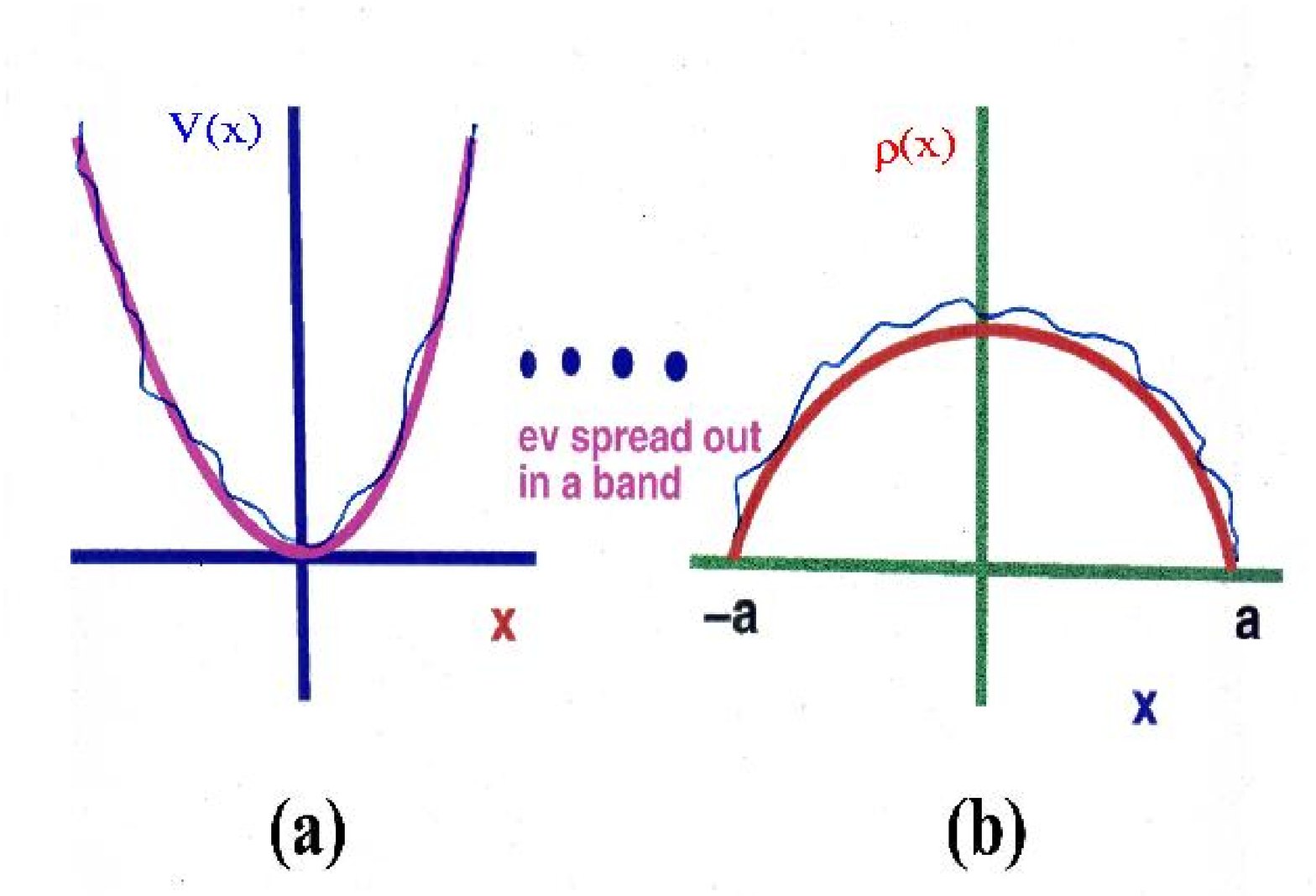}


\caption{ (a). The confining potential (b). The density of eigenvalues }
\label{fig1ab} 
\end{figure}

This is a system of $N$ particles with coordinates $\lambda_i$ on
the real line, confined by the potential $V$ and repelling each 
other with a logarithmic repulsion. The
spectrum or the density of eigenvalues $\rho (x) =
\langle {1\over N} \sum_{i=1}^N \delta (x-\lambda_i) \rangle $ 
is, in the large N limit, just the Wigner semi-circle for a quadratic
potential. The physical picture is that the eigenvalues try to be at the 
bottom of the well. But it costs energy to sit on top of each other
because of logarithmic repulsion, so they spread. $\rho$ has support on a 
finite line segment. This continues to be true whether the potential is
quadratic or a more general polynomial and only depends on there being
a single well though the shape of the Wigner semi-circle is correspondingly 
modified. For the quadratic potential $V(x)={\mu\over 2} x^2$
the density is 
$\rho (x) = {1\over \pi} \sqrt{(x^2-a^2)}$ where $x \in [-a,a]$,
and $\rho(x)=0$ elsewhere. The region $[-a,a]$ is said to be the `cut'
where $\rho$ has support. The end of the cut is given by 
$a=\sqrt{2\over \mu}$. See Fig. \ref{fig1ab}.

On changing the potential more drastically by having two wells
the density can get a support on two disconnected segments 
ref. \cite{shima82,ddjt}.
The simplest example is the potential 
$V (M) = {\mu\over 2} M^2 + {g\over 4} M^4$ with $\mu < 0, 
g > 0$. When the wells are
sufficiently deep, specifically, when $|\mu|> 2\sqrt{g}$, 
the density of eigenvalues is given by

\beql
\rho (x) &=& {g\over \pi} x \sqrt{(x^2-a^2)(b^2-x^2)},  ~~~~~~~~   
a\leq x \leq b ~ or ~ -b\leq x \leq -a\nonumber\\   
&=& 0 ~~~~~~~~ otherwise
\label{rxd}
\eeql
\noindent
where $ a^2 = {1\over g} [|\mu|-2\sqrt{g}]$ and
$ b^2 = {1\over g} [|\mu|+2\sqrt{g}]$. 
The eigenvalues sit in symmetric bands
centered around each well. Thus $\rho$ has support on two line
segments. As $|\mu|$ approaches $2\sqrt{g}$ from below, 
$ a \rightarrow 0$ and the two bands merge at the origin. 
For $\mu > - 2 \sqrt{g}$, the density is 
\beql
\rho (x) &=& {{g x^2}\over \pi} \sqrt{x^2-{2\mu\over g}}
~~~~~~~~ -\sqrt{2|\mu|\over g} < x < \sqrt{2|\mu|\over g} \nonumber \\
&=& 0 ~~~~~~~~ otherwise.
\label{rxs}
\eeql

\begin{figure} 

\leavevmode
\epsfxsize=4in
\epsffile{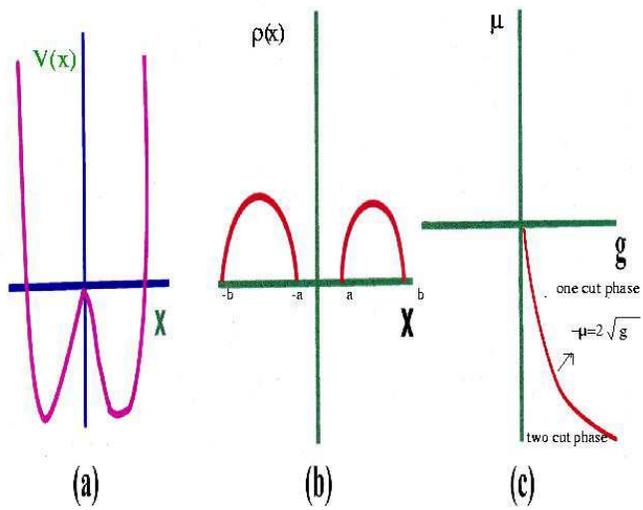}

\caption{ (a). The double-well potential (b). Density of eigenvalues 
(c). The phase diagram }
\label{fig2ab} 
\end{figure}

The phase diagram and density of eigenvalues for this case 
is shown in Fig. \ref{fig2ab}.

The simplest way to determine $\rho (z)$ explicitly is to use the
generating function $F(z)=\langle {1\over N} Tr {1\over {z-M}} \rangle$. 
$F(z)$ satisfies a Schwinger-Dyson equation whose solution is
$ F(z) = {1\over 2} [ V^{\prime} (z) + \sqrt{\Delta} ] $
with $\Delta (z) = V^{\prime} (z)^2 - 4 b(z)$
and $b(z) = g z^2 + \mu + g \langle {1\over N} Tr M^2 \rangle$ 
(see ref. \cite{bdjt}). The density
$\rho (x)$ is then determined by the formula 
$\rho (z) = -{1\over 2\pi} Im \sqrt{\Delta (z)}$. 
In what follows I will outline the recurrence coefficient method 
of the orthogonal polynomials to establish that there exist
multiple solutions which give the same free energy $\Gamma$, $F(z)$ and
$\rho (z)$ in the large $N$ limit but differ at higher orders. 
Then I give the results for the 
two-point correlators (also known as the 
`smoothed' or `long range' correlators); these tend to
different limits as $N$ is taken to infinity along different sequences
(odd or even). This property is similar to that suggested in
another model of glasses ref. \cite{mpr94}.
   
\section{Orthogonal Polynomial Approach}
\label{ortho}

The partition function $Z$ can be rewritten
in terms of the orthogonal polynomials 
$P_n$. These are defined as  $P_n = \lambda^n + 
C^{(n)}_{n-1}\lambda^{n-1}+\ldots + C^{(n)}_1 \lambda 
+ C^{(n)}_0$, where $C^{(n)}_i$ are constants, and satisfy
the orthogonality conditions
$ (P_n,P_m) = \int_{-\infty}^{\infty} d\lambda e^{-NV(\lambda)}
P_n (\lambda) P_m (\lambda) = h_n \delta_{nm}
$ . For example: $P_0 (\lambda)=1$,
$P_1 (\lambda) = \lambda + c^{(1)}_1$, $P_2 (\lambda) 
= \lambda^2 + c^{(2)}_1 \lambda + c^{(2)}_0 \ldots$. Then the
partition function in terms of the orthogonal polynomials is, 
ref. \cite{biz},
\beq
Z = \int \prod_{i=1}^N d\lambda_i e^{-N \sum V(\lambda)}
\left|
\begin{array}{ccc}
P_0 (\lambda_1)&....&P_0 (\lambda_N)\\
P_1 (\lambda_1)&....&P_1 (\lambda_N)\\
.\\
.\\ 
.\\
P_{N-1} (\lambda_1)&......&P_{N-1} (\lambda_N)\\
\end{array}
\right|.
\eeq
$Z$ can be expressed in terms of the $h_n$'s; 
$
Z=N! h_0 h_1 h_2.....h_{N-1}.
$
For example: in the $N=2$ case,
$
Z=\int d\lambda_1 d\lambda_2 e^{-N V(\lambda_1)-N V(\lambda_2)}
(P_0 (\lambda_1) P_1(\lambda_2)-P_0 (\lambda_2) P_1 (\lambda_1))^2
= h_0 h_1 + h_0 h_1  
= 2! h_0 h_1.
$ 

\subsection{The Recurrence Coefficients}\label{reccoef}

The $P_n$ satisfy recurrence relations, ref. \cite{biz},
\beq
x P_n = P_{n+1} + S_n P_n + R_n P_{n-1}
\label{rec}
\eeq
where $R_n$ and $S_n$ are the recurrence coefficients which depend
on the potential. These recurrence coefficients are central to our analysis,
since the free energy and all correlation functions can be expressed in terms
of them. The reason why the $P_n$'s satisfy such a simple 
recursion equation is that $ \int x P_n P_{n-2} e^{-NV(x)} dx =0 $
because $ x P_{n-2} $ is a polynomial of degree $n-1$ and can be expressed
as a superposition of $P_{n-1},P_{n-2}, \ldots P_0$. 
Thus $P_{n-2}$ and lower order polynomials 
do not appear on the right hand side of the recurrence relation
eq. (\ref{rec}). It can be shown that $h_n=h_{n-1}R_n$. Thus
the product
$
h_0 h_1 ...... h_{N-1} = 
h_0 (h_0 R_1) (h_0 R_1 R_2) ...... (h_0 R_1... R_N) 
= h_0^N R_1^{N-1} R_2^{N-2} ...... R_{N-1}
$, hence the free energy  
$
\Gamma = \ln Z = \ln N! + N \ln h_0 + \sum_{n=1}^{N-1} (N-n) \ln R_n.
$

To solve for the recurrence coefficients we have three methods 
based on (i). Integrals, (ii). Recurrence Relations, and 
(iii). an Effective Potential.

\subsubsection{Integrals}

The recurrence coefficients can be determined by integrals or
moments $I_n= \int dx x^n e^{-NV(x)}$ as described below. 
The $P_n$'s can be expressed in terms of $R_n$ and $S_n$ as 
$P_{n+1} = x P_n - S_n P_n - R_n P_{n-1}$,
with $P_0 = 1$, $P_1 = (x-S_0)$ and $ P_2 = (x-S_1)(x-S_0)-R_1 $ etc. 
Consider the equation
$
0 = (P_0,P_1) = \int dx (x-S_0) e^{-N V(x)} 
$,
this results in an integral equation for the coefficient $S_0$ i.e.
$
S_0 = {1\over h_0} \int dx x e^{-N V(x)} = {I_1\over I_0}.
$
As another example
the recurrence coefficient $R_1$ can be determined as follows.
The integral expression for the recurrence coefficient $R_1$ is then found
from $R_1={h_1\over h_0}$ where
$h_0 = \int dx e^{-N V(x)} = I_0$
and $h_1=\int dx e^{-NV(x)} P_1^2 (x)
= \int dx e^{-NV(x)}(x-S_0)^2= 
\int dx e^{-NV(x)}(x^2-2xS_0+S_0^2)=I_2-2I_1S_0+S_0^2$.
Hence $R_1= {I_2\over I_0} - ({I_1\over I_0})^2$. Similar 
expressions for all the other recurrence coefficients in 
terms of integrals can be found. 

\subsubsection{Recurrence Relations}

The recurrence coefficients satisfy recurrence relations that 
follow from the identities \cite{biz}

\beq
\noindent
I. ~~\int dx e^{-NV(x)} P_n (x) V^{\prime} (x) P_n (x) =0
\eeq

and

\beq
\noindent
II. ~~ n h_{n-1} = N \int dx e^{-NV(x)} P_n (x) 
V^{\prime} (x) P_{n-1} (x)
\eeq

I follows from the identity
$0=\int P_n (x) P^{\prime}_n (x) e^{-NV(x)}$
which holds because $P_n^{\prime}$, being a linear combination
of $P_{n-1}$ and lower order polynomials, is orthogonal to $P_n$.
II follows from the identity 
$ \int P^{\prime}_n (x) P_{n-1} (x) e^{-NV(x)}=n h_{n-1}$. 

Let us take the following examples:

\noindent
(a) $V(x)={\mu\over 2} x^2$.\\ 
For this potential the recurrence 
relation I is $0=\mu(P_n,P_{n+1}+S_nP_n+R_nP_{n-1})=S_nh_n$, which implies
$S_n=0$, while the recurrence relation II is 
$n h_{n-1} = \mu N (P_n,P_n+S_{n-1}P_{n-1}+R_{n-1}P_{n-2})    
= \mu N h_n$ which implies ${n\over N} = \mu R_n $.
This determines exactly the recurrence coefficients $S_n=0$
and $R_n={n\over {N \mu}}$ for this potential.

\noindent
(b) $V(x)={\mu\over 2} x^2 + {g\over 4} x^4$. \\ 
The recurrence relation I is 
\beq
0=\mu S_n + g [ R_{n+1} (S_{n+1}+2S_n)+ R_n (2S_n+S_{n-1}) + S_n^3]
\eeq
while II is 
\beq
{n \over N} = \mu R_n + g (R_{n-1}+R_n+R_{n+1}) + S_n^2 + S_{n-1}^2
+S_{n-1}S_n.
\eeq

Using the initial values for $R_0=0$, $S_0$ and $R_1$ we can determine
$S_1,R_2 \ldots$ using the above recurrence relations for the recurrence
coefficients. 

For this potential we now consider the two cases corresponding to $V$ 
having one or two wells.

\begin{figure} 

\leavevmode
\epsfxsize=4in
\epsffile{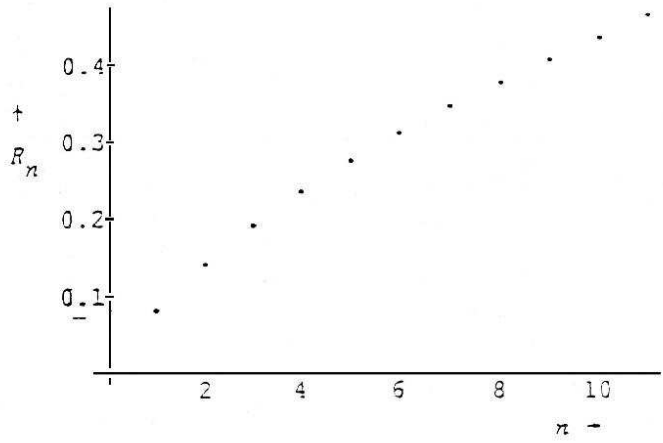}


\caption{Recurrence coefficients for the single well}
\label{fig3} 
\end{figure}

\noindent
(1) The 1-well case: $\mu>0$, $g>0$. \\
It follows that $S_n=0$, because
all $I_n$ are zero for odd $n$ whenever $V(x)$ is an even function of 
$x$. Then the recurrence relation I is trivially satisfied. 
The recurrence relation II is
\beq
{n\over N} = R_n (\mu + g (R_{n+1}+R_n+R_{n-1})).
\label{recr}
\eeq
Thus $R_{n+1}={n\over {g N R_n }} - {\mu\over g} - R_n - R_{n-1}$.
We determine $R_1={I_2\over I_0}-({I_1\over I_0})^2$ by evaluating 
$I_2,I_1$ and $I_0$ 
numerically and evaluate $R_n$ for $n > 1$ using this equation.  
The result, shown in Fig. \ref{fig3}, suggests that $R_n$'s lie on 
a smooth curve. This curve is analytically determined as follows: 
For the large $N$ limit we set
$ {n\over N} = x $ and make the ansatz that $R_n$ is a smooth function
of $x$ and expand as  
$R_n = R(x) + {1\over N} R_1 (x) + {1\over N^2} R_2 (x) + \ldots$ .
Then to leading order in $N$ eq. (\ref{recr}) implies that 
$ x=R(x)[\mu+3g R(x)] $
which is a quadratic equation in $R(x)$ with the solution
$ R(x)= {1\over 6g} [-\mu \pm \sqrt{\mu^2+12gx}] $.
This fits very well with the numerical evaluation of $R_n$ 
which can be approximated by a smooth curve at large N as shown
in Fig. \ref{fig3}.

The generating 
function for the single well case is given by (see ref. \cite{bdjt})
\beq
F(z) = \int_0^1 dx {1\over \sqrt{z^2-4 R(x)}}.
\label{fzr}
\eeq
On substituting $R(x)$ in the above equation for the generating
function one gets the same answer as that 
found by the Schwinger-Dyson equation. Eq. (\ref{fzr}) yields the
expression eq. (\ref{rxs}) for the density of eigenvalues, with a
single cut.

(2). The 2-well case, $\mu<0$, $g>0$. \\
Fig. \ref{fig4} exhibits a numerical result for $R_n$ in this region 
of phase space which shows that the assumption
of a single smooth function describing $R_n$ is no longer correct.
It suggests the following ansatz for $R_n$ at large $N$:

\begin{figure} 

\leavevmode
\epsfxsize=4in
\epsffile{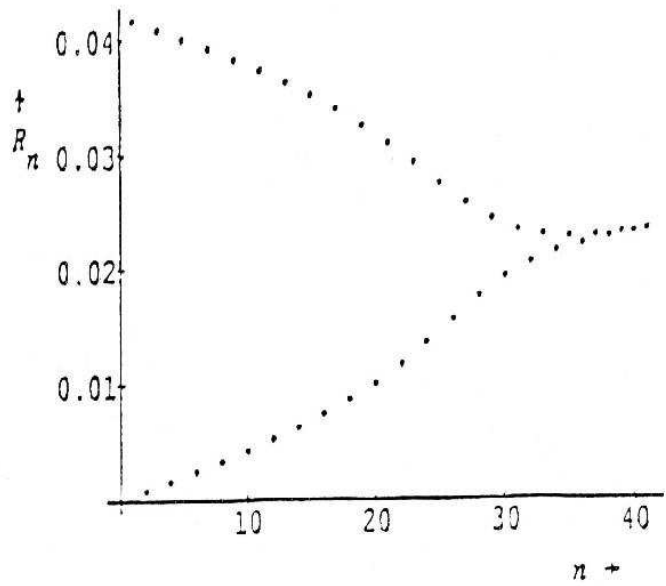}


\caption{Recurrence coefficients for the double well}
\label{fig4} 
\end{figure} 

For $n\leq \bar{n}$ ( for some $\bar{n}$ to be determined), we have
a `period-2' structure:
\beql
R_n = \left\{ 
\begin{array}{ll}
A_n 
= A(x)+{1\over N} A_1 (x)+\ldots & 
for ~~ n ~~ even \\
B_n  
= B(x)+{1\over N} B_1 (x)+\ldots &
for ~~ n ~~ odd, 
\end{array}
\right.
\label{recab}
\eeql
and for $n > \bar{n}$, we have a `period-1' structure (as for the
one-well case):
\beq
R_n = R(x)+{1\over N} R_1 (x) + \ldots 
\label{rone}
\eeq
We denote $\bar{x}={ {\bar n}\over N}$.
Substituting eq. (\ref{recab}) into eq. (\ref{recr}) and equating
equal powers of $1\over N$ we get, for $x\leq \bar{x}={\mu^2\over 4g}$,
\beql
A(x) &=& {1\over 2g} [ |\mu| - \sqrt{\mu^2-4gx} ], \nonumber \\
B(x) &=& {1\over 2g} [ |\mu| + \sqrt{\mu^2-4gx} ]. 
\label{ab}
\eeql
For $x\geq \bar{x}$, substituting eq. (\ref{rone})
into eq. (\ref{recr}) we get
\beql
R(x) &=& {1\over 6g} [ -\mu + \sqrt{\mu^2+12gx} ].
\label{r}
\eeql
The above analytical result and numerical figure agree very well.
Note that $\bar{x}=1$ is the equation for the phase boundary between
1-cut and 2-cut phases (see section 2) in the double well region of
parameter space.
The generating function in terms of the recurrence
coefficients for the two well case is ( see ref. \cite{bdjt}
for a derivation )
\beql
F(z) = \int_0^1 dx {z\over \sqrt{(z^2-(A(x)+B(x)))^2-4AB}}, 
~ &&{\rm if} ~ \bar{x}>1, \nonumber \\
= \int_0^{\bar{x}} dx {z\over \sqrt{(z^2-(A+B))^2-4AB}} 
+ \int_{\bar{x}}^1 dx {1\over \sqrt{x^2-4R(x)}} 
~ &&{\rm if} ~ \bar{x}<1. \nonumber \\  
\label{f}
\eeql
When $\bar{x}>1$, i.e. in the 2 cut phase,
substituting $A,B$ from eq. (\ref{ab}), $ F(z)={1\over 2} 
[\mu z + g z^3 - gz \sqrt{(z^2-a^2)(z^2-b^2)}]$
which is the same as obtained by the Schwinger-Dyson equation.
This yields the solution eq. (\ref{rxd})for the density of eigenvalues. 
When $\bar{x}<1$, eqs. (\ref{ab}) and (\ref{r}) leads to the one cut 
solution eq. (\ref{rxs}).

\subsubsection{Effective potential method for determining recurrence
coefficients}

\begin{figure} 

\leavevmode
\epsfxsize=5in
\epsffile{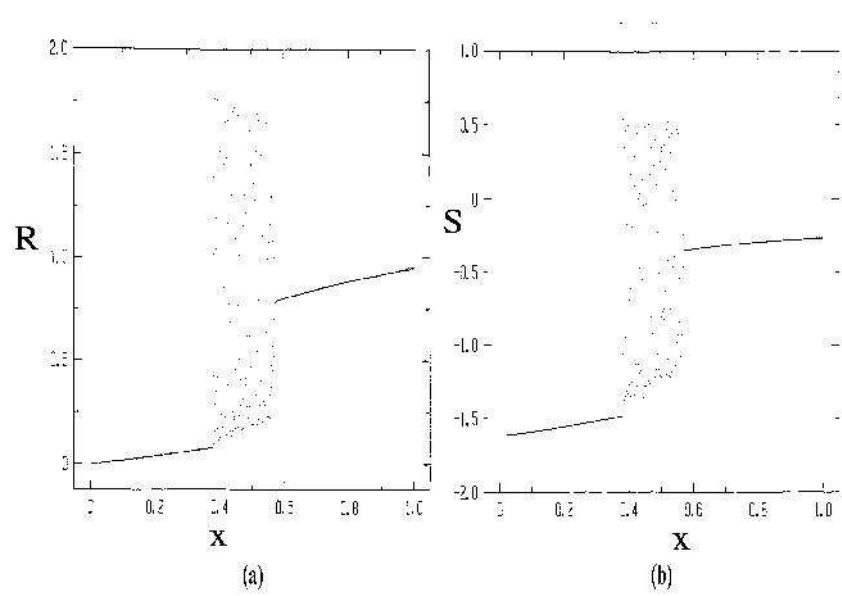}


\caption{(a). Recurrence coefficients $R_n$ for the asymmetric double well
(b). Recurrence coefficients $S_n$ for the asymmetric double well}
\label{fig5ab} 
\end{figure}

\begin{figure} 

\leavevmode
\epsfxsize=4in
\epsffile{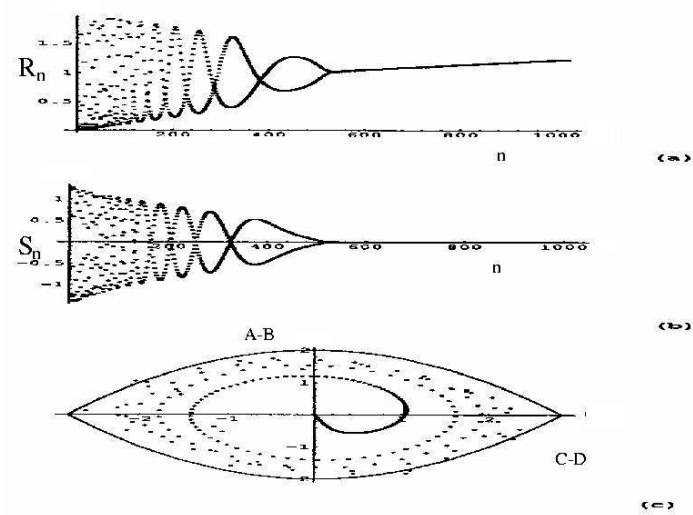}


\caption{ Graphs of recursion coefficients for a spontaneously broken 
solution of the double-well potential with $N=512, \mu=-2, g=1$  
(a). Recurrence coefficients $R_n$ and
(b). Recurrence coefficients $S_n$ after 100000 minimization steps
from a random start.
(c). Orbit in the A-B vs C-D plane.}
\label{fig6abc} 
\end{figure}

Numerically we can evaluate the recurrence coefficient by minimizing an
effective potential $V_{eff}$ that can be determined from the recurrence
relations e.g. for $V(x) = \sigma x + {\mu \over 2} x^2 + {g\over 2} x^4$, 
the recurrence relations are:
\beql
{n\over N} &=& R_n [ \mu + g (R_{n+1}+R_n+R_{n-1}
+S_n^2+S_{n-1}^2+S_{n-1}S_n) ], \nonumber \\
0 &=& \sigma + \mu S_n + g [ R_{n+1}(S_{n+1}+2S_n)+R_n(2S_n+S_{n-1})+S_n^3 ].
\label{rssig}
\eeql
It is easy to see that if one defines
\beql
V_{eff} &=& \sum_{n=0}^{\infty} \{ {-n\over N} ln R_n + \mu R_n \nonumber \\
&+& {g\over 2} (R_n^2 + 2 R_n R_{n+1}) + \sigma S_n + {\mu \over 2} S_n^2
\nonumber \\
&+& {g\over 4} S_n^4 + g R_n (S_n^2 + S_{n-1}^2 + S_{n-1} S_n) \},
\label{veff}
\eeql
then the recurrence relations follow by setting ${{\partial V_{eff}} \over
{\partial R_n}}=0$ and ${{\partial V_{eff}} \over
{\partial S_n}}=0$. For $\sigma=0$, setting $S_n\equiv 0$ in eq. (\ref{veff})
and minimizing with respect to $R_n$ yields period 1 or period 2 solutions
for $R_n$ as shown in figures 3 and 4. However, when $\sigma\neq 0$, 
$\mu<0, g>0$, (i.e. $V(x)$ has two wells which are asymmetric), then
numerical minimization of eq. (\ref{veff}) with respect to $R_n$ and $S_n$
yields a solution of the type displayed in Fig. \ref{fig5ab}. 
The figure suggests that the recursion 
coefficients possibly become chaotic. However, whether they are really 
chaotic or whether this is only apparently so, requires more detailed
numerical work \cite{lrr}. It would be rather interesting to
do this and characterize this chaos and also
to understand why the recursion coefficients are so complicated.
At any rate, this suggests that it might be interesting to explore
solutions of the eq.'s (\ref{rssig}) for non-zero but small $\sigma$, and
see whether at $\sigma=0$, there are alternate solutions ( other than the
one with $S_n\equiv 0$ discussed earlier ). This indeed turns out to be the
case. Fig. \ref{fig6abc} shows a numerical solution to eq. (\ref{rssig})
with $\sigma=0$, obtained by first obtaining solutions for progressively
decreasing nonzero values of $\sigma$, and using the solution for a particular
value of $\sigma$ as an initial condition to get a solution for the next 
smaller value of $\sigma$. As seen in Fig. \ref{fig6abc}, even at $\sigma=0$,
we can get a solution with $S_n\neq 0$ by this procedure. As we show 
analytically in the next sections, there is a large family of multiple 
solutions that exist at $\sigma=0$. One possible explanation for the apparent 
chaos in Fig. \ref{fig5ab} is that a nontrivial mixing between multiple
solutions might be occurring.

\section{Multiple Solutions In The Symmetric $M^4$ 
Matrix Model}\label{multi}

The simplest way to understand the existence of 
multiple solutions which appear in
these models is to consider the following integral 

\beq
S_0 = {I_1\over I_0} = { { \int_{-\infty}^{\infty} dx x e^{-N V(x)} } 
\over { \int_{-\infty}^{\infty} dx e^{-N V(x)} } }
\label{so}
\eeq

for $V(x)= {\sigma} x + {\mu\over 2} x^2 + {g\over 4} x^4$ 
with $\mu < 0$ and $g > 0$, in the limit $\sigma\rightarrow 0$, 
$V$ has a maximum at $x=0$ and 
minimum at $x=x^{\pm}=\pm\sqrt{|\mu|\over g}$.
The integral $I_1$ in eq. (\ref{so})  
is zero if we take $\sigma\equiv 0$, and then 
take $N \rightarrow \infty$ for then the integrand
is an odd function. Hence $S_0=0$ for $\sigma=0$.
However, let us take $\sigma \ne 0$ first, evaluate the integral, in 
the limit $N \rightarrow \infty$ and then take the limit 
$\sigma \rightarrow 0$. The integral can be evaluated using the
saddle point method. Since for $\sigma\neq 0$, one of the minimas
gives dominant contribution, the result is $S_0=x^+$ as 
$\sigma\rightarrow 0^-$. Thus $S_0$ depends upon the order of
the limits $N \rightarrow \infty$ and $\sigma \rightarrow 0$.

A more precise way of establishing the presence of multiple solutions
in these models is by using the recurrence coefficients; the previous
sections have built up the tools that will be used here.
Let us relax the condition $S_n = 0$; we will find that 
this will be interesting. Consider a period two ansatz for 
both $R_n$ and $S_n$. Then 
\beql
R_n &\rightarrow& A_n, ~~ S_n \rightarrow C_n, {\rm for~~n~~even,~~and} 
\nonumber\\   
R_n &\rightarrow& B_n, ~~ S_n \rightarrow D_n, {\rm for~~n~~odd}.
\eeql
As before in the large $N$ limit setting $A_n$, $B_n$, $C_n$ and $D_n$
to be equal to smooth functions of $x\equiv {n\over N}$,
denoted $A$, $B$, $C$, $D$ respectively, one finds that
eq.'s (\ref{rssig}) with $\sigma=0$ reduce to
\beql
n~~ even ~~ x &=& A[\mu+g(2B+A+C^2+D^2+CD)] \nonumber \\ 
n~~ odd~~  x &=& B[\mu+g(2A+B+C^2+D^2+CD)] \nonumber \\
n~~ even~~ 0 &=& \mu C + g (B(D+2C)+A(D+2C)+C^3) \nonumber \\
n~~ odd~~ 0 &=& \mu D + g (A(C+2D)+B(C+2D)+D^3).
\eeql 
There are four equations and four unknows here 
but after some work we find that there is only 
three independent equations. The three independent equations are
\beql
C+D &=& 0 \nonumber \\
A+B+C^2 &=& {-\mu\over g}\nonumber \\
AB&=&{x\over g}.
\label{abcd}
\eeql
Thus there is a infinite class of solutions labelled by one function of
$x$ in the large $N$ limit.
For example, let us consider the two extreme solutions. (i). The `symmetric
solution', characterized by $C-D=0$. Then eq. (\ref{abcd}) implies 
$C=D=0$, and $A(x)$, $B(x)$ are
given by eq. (\ref{recab}). This is the same solution as discussed
in section 3. (ii). The `maximally asymmetric solution' characterized by,
$A-B=0$. Then eq. (\ref{abcd}) implies
$A=B=R=\sqrt{x\over g}$ and
$C=-D=[{|\mu|\over g}-\sqrt{4x\over g}]^{1\over 2}$ .
The entire infinite class of solutions have the same  
eigenvalue density and free energy in the large $N$ limit.
This can be seen by evaluating the generating function which
turns out to be 
\beq
F(z) = \int_0^1 dx { {2z-(C+D)}\over {\sqrt{(z^2-z(C+D)-(A+B-CD))^2-4AB}} }. 
\label{fz}
\eeq

Eq. (\ref{fz}) contains precisely the same three combinations that are fixed 
by the recurrence relation eq. (\ref{abcd}). 
Therefore independent of which solution is
chosen we get the same $F(z)$. Since $F(z)$ determines $\rho$
and the latter determines $\Gamma$ at large $N$, this proves
that in the limit $N\rightarrow \infty$ we have an infinite set
of solutions of the recurrence relations with the same eigenvalue
density and free energy. This
demonstrates the presence of multiple solutions from the
recurrence coefficient point of view. 

\section{Non-perturbative Solution}\label{nonpert}

The multiple solutions found above show differences at higher
powers of $1\over N$, for example, in the double-scaling limit
ref. \cite{ddjt,dss,cm,ms,nappi,pet,cdm,hmpn} which I describe in this section. 
Let us begin by taking the symmetric solution and proceed
by expanding the even, odd recurrence 
coefficients $A, B$ as
\beql
A_n &=& a_0 + \epsilon (f_e (t) + f_0 (t)) 
+ \epsilon^2 (r_e (t) + r_0 (t)) ~~n ~~even \nonumber \\
B_n &=& a_0 + \epsilon (f_e (t) - f_0 (t))
+\epsilon^2 (r_e (t) - r_0 (t)) ~~n ~~odd 
\eeql
where $x={n\over N}=1-\epsilon^2 t$, $a_0={(-\mu)\over 2g}$, and
$\epsilon=N^{-1/3}$. Then upon
equating equal powers of $\epsilon$, we get
$f_e=r_0=0$, $f_0=f(t)$, $r_e = { {f^2-t}\over{4}}$, where the 
susceptibility is given by $\chi \approx { {f^2+t} \over 4}$. 
$f$ satisfies $f''-{1\over 4} f^3 + {1\over 2} ft = 0$ the 
Painleve II equation. 

We then take the asymmetric solution with
$ A-B \ne 0 $, $ C-D \ne 0$. On making the following 
expansion
\beql
C_n &=& \epsilon g(t) + \epsilon^2 \ldots\nonumber \\
D_n &=& -\epsilon g(t) + \epsilon^2 \ldots
\eeql
we get the following coupled equation ref. \cite{nappi,pet,cdm,hmpn}
$
f''- {1\over 4} f (g^2+f^2) + {1\over 2} ft = 0$ and
$
g''- {1\over 4} g (g^2+f^2) + {1\over 2} gt = 0. 
$
If we make the substitution 
$ f = r \cos \theta$, $g = r \sin \theta$ where $r$ and $\theta$
are functions of $t$, then $r$ satisfies a modified Painleve equation
\beq
\ddot{r} - {1\over 4}r^3 + {1\over 2} t r - {l^3\over r^3} = 0,
~~
r^2 \dot{\theta} = l = const.
\eeq
The case $l=0$ is the Painleve II equation. The susceptibility
$\chi \approx {{f^2+g^2+t}\over 4} = {{r^2+t}\over 4}$ and
$\chi \approx {3t\over 4}-({{1+l^2}\over 4}) t^{-2}+\ldots$
which is very different from the symmetric case. 
Thus the multiple solutions show
differences at higher order in $1\over N$.    

\section{Distinguishing multiple solutions: Correlators, Odd and Even N}
\label{corr}

Multiple solutions are distinguished by certain correlators.
Consider the correlator $ \langle Tr M ~~ Tr M \rangle_c = 
\langle Tr M ~~ Tr M \rangle
-\langle Tr M \rangle^2 $ where $\langle O \rangle =
{1\over Z} \int dM e^{-N Tr V(M)} O$.
It can be shown \cite{bdjt} that 
$\langle Tr M ~~ Tr M \rangle_c=R_N$.
Since $R_N$ depends on the choice of the solution, this 
demonstrates that the multiple solutions under discussion
give rise to different correlation functions in general. In
particular note that $R_N=A_N$ if $N$ is even and $R_N=B_N$
if $N$ is odd. For the `symmetric solution' $A\neq B$, hence 
this correlator changes by $O(1)$ as $N$ goes from odd to even.
However for the maximally asymmetric solution $A=B$, hence this 
correlator remains essentially the same as $N$ is changed from
odd to even. 

Another example of a correlator which distinguishes between the two 
solutions is 
\beq
\langle Tr M ~~ Tr M ~~ Tr M \rangle_c = R_N (S_{N-1}-S_N).
\eeq
For the symmetric solution $S_n=0$ for all $n$, hence this
correlator vanishes. For the asymmetric solution $S_n$ is 
period two and the correlator changes sign while going from 
odd to even $N$. 

\begin{figure} 

\leavevmode
\epsfxsize=4in
\epsffile{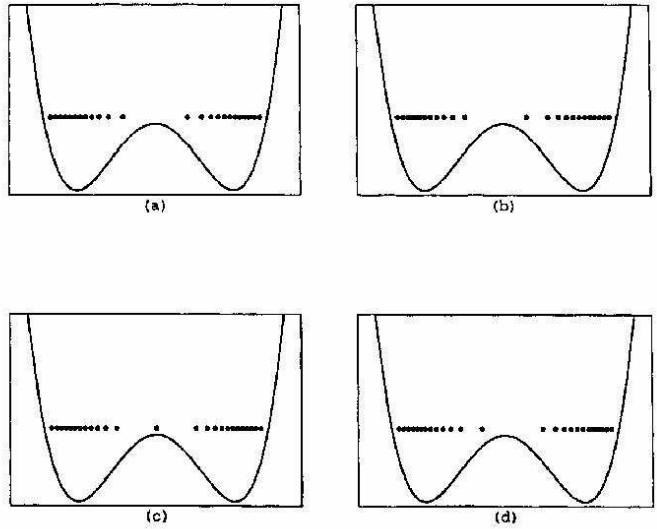}


\caption{Eigenvalue distribution for even and odd N: (a) N=24,
symmetric solution; (b) N=24, asymmetric solution; (c) N=25,
symmetric solution; (d) N=25, asymmetric solution.}
\label{fig7abcd} 
\end{figure}

The orthogonal polynomial $P_n$ can be thought of as a 
`wavefunction' of a `state' $|n \rangle$ in the `coordinate basis',
i.e. $\langle \lambda | n \rangle = {P_n(\lambda)\over \sqrt{h_n}} 
e^{-{N\over 2}V(\lambda)}$, where $|\lambda\rangle$ are eigenstates 
of the operator $\hat{M}$. The operator $\hat{M}$ is defined in 
the basis of states $|n\rangle$ as the matrix

\beq
\left[
\begin{array}{ccccc}
S_0& \sqrt{R_1}& 0 & \ldots \\
\sqrt{R_1}&S_1&\sqrt{R_2} & \ldots \\
0&\sqrt{R_2}&S_2& \ldots \\
\end{array}
\right],
\eeq

\noindent
which follows from the recurrence relation satisfied by the 
orthogonal polynomials eq. (\ref{recr}); see ref. \cite{kgm}
for more details. It can be shown that the density of eigenvalues
of this matrix in the large $N$ limit is precisely $\rho(\lambda)$.
It is therefore of interest to calculate the eigenvalues of this
matrix. Since the matrix elements are given by the recurrence 
coefficients we can determine the eigenvalues
numerically (see section (\ref{reccoef}))
from a given solution of the recurrence coefficients. 
The figures show the location of the 
eigenvalues for the symmetric solution and the maximally
asymmetric solution (see Fig. (\ref{fig7abcd})).

For $N$ even, half the eigenvalues are in one well and the other half
in the other well. While this is true for both solutions the detailed 
positions are not the same. When $N$ is odd, for the maximally
asymmetric solution 
one extra eigenvalue is located in one of the wells (the well it is 
located in depends on the sign of $S_0$). 
For the symmetric solution this extra eigenvalue is in the 
center, thus preserving the symmetry between both wells. This seems 
to suggest that the bulk effect of the two solutions is the same but 
they differ by one eigenvalue effects.

Further evidence for the effects of multiple solutions comes from
correlation functions of the density operator $\hat{\rho} (x)=
{1\over N} Tr \delta (x-M)$. The explicit expression for the smoothed 
two-point correlator for a general 2-cut solution, obtained by the 
method of steepest descent \cite{d,bd} is

\beql
4 \pi^2 N^2 <\hat{\rho} (\lambda) \hat{\rho} (\mu)>_c = 
{ {\epsilon_{\lambda}\epsilon_{\mu}} \over { \beta \sqrt{|\sigma(\lambda)|}
\sqrt{|\sigma(\mu)|} } } ~~~~~~~~~~ \nonumber \\
\left( { {\sigma(\lambda)+\sigma(\mu)}\over {(\lambda-\mu)^2} } 
+ { {\sigma^{\prime} (\lambda) - \sigma^{\prime} (\mu)}
\over {(\lambda-\mu)} } +  
\lambda^2 + \mu^2 -{s\over 2} (\lambda+\mu) + 2C \right).
\label{rr}
\eeql

Here $\sigma(z)=\prod_{i=1}^4 (z-a_i)$, $s=a_1+a_2+a_3+a_4$, (the
support of the eigenvalues consists of the two segments $[a_1,a_2]$
and $[a_3,a_4]$),
$\epsilon_{\lambda}=+1~~ {\rm for}~~ \lambda \in [a_3,a_4], 
\epsilon_{\lambda}=-1~~{\rm for}~~\lambda \in [a_1,a_2]$ and 
$\beta=1,2,4$ depending on whether $M$ the matrix
is real orthogonal, hermitian or self-dual quartonian. 
$C$ is an undetermined constant in this method. It turns out that the
same correlator can be calculated using other methods, which yield 
different values of $C$. For example in \cite{bd} we obtained an 
asymptotic form of $P_n(x)$ for the symmetric double well potentials 
for $n$ close
to $N$. Using this form of $P_n$ (which corresponds to the 
`symmetric solution'), we found the smoothed $<\hat{\rho}\hat{\rho}>_c$
given by eq. (\ref{rr}), with $C=(-1)^N$, (the origin of this nontrivial
$N$ dependence, also observed in \cite{kf}, is explained in 
\cite{david}). On the other hand in ref. \cite{aa}, this correlator
was calculated by the loop equation method starting from an asymmetric 
$V$ and taking the limit of a symmetric potential where $C$ was found to
be $-{1\over 2}\{ (a^2+b^2)-(a+b)^2[{E(k)\over K(k)}]\}$ where $E(k)$
and $K(k)$ are complete elliptic integrals of the first and second kind
and $k={{2\sqrt{ab}}\over {(a+b)}}$. We conjecture 
that different values of $C$ corresond to different solutions of the 
recurrence coefficients. 

In this section, we have shown the multiple solutions discussed
in section 4 give rise to different correlation functions. An 
interesting feature of the solutions is that there is a nontrivial $N$
dependence which survives in the large $N$ limit. In particular,
for the symmetric solution we have shown a difference between
odd and even $N$ in the thermodynamic limit. This is reminiscent of
the behaviour in a model for glasses \cite{mpr94}. 

\section{The connection of random matrix models with a simple
model of structural glasses}\label{glass}

This section is concerned with building up a connection
with the study of structural glasses in the high temperature
phase, starting from the work of ref. \cite{ckpr,p}. 
The hamiltonian
$
H = {1\over n} \sum_{a\ne b} (s^a.s^b)^p
$
corresponds to $n$ particles moving in an $N$-dimensional space, 
in the limit $n, N \rightarrow \infty$. The coordinates of the
particles are $s^a=(s^a_1 ...... s^a_N)$ and $a=1......n$. 
In the `spherical' case the particles are constrained to move 
on the surface $|s^a|^2=N$, for all $a$. In the `Ising' case they 
occupy only the vertices of a hypercube $s^a_i=\pm 1$. 

The corresponding partition function is
$
Z= e^{\beta N^2} Tr_{\{s\}} exp[-{\beta\over {\alpha N}}
tr (S^{\dagger}SS^{\dagger}S)]
$
where $S$ is the $N\times n$ rectangular matrix of elements
$s_i^a$, and $Tr_{\{s\}}$ runs over either the spherical or the 
Ising measure. Here $n=\alpha N$ with $\alpha \ge 1$. Using
the large $N$ equivalence with the global constraint one gets

\beq
Z_{sph} \approx e^{\beta N^2} \int d\mu \int \prod_{i=1}^N 
dx_i exp \left( -N^2 {{\beta \mu}\over 2} - \beta E [x]
\right)
\eeq

The $x_i \sqrt{N}$ are the ``diagonal'' values of $S$ in its 
canonical form and $E[x]\equiv N \sum_{i=1}^N V(x_i) - 
{1\over 2 \beta} \sum_{i\ne j} ln|x_i^2-x_j^2|$ 
where $x=(x_1,x_2,...,x_N)$ and $V(x)$ is given by
\beq
V(x)=\left( {1\over \alpha} x^4 - 
{\mu \over {2\alpha}} x^2 - {(\alpha-1)\over \beta}\ln|x|\right).
\label{penner}
\eeq
This potential has the form of the generalized
Penner model. I show 
briefly (see ref. \cite{bdjt}) that a very closely related model
that can be exactly solved, the Gaussian
Penner model, has 2 cuts and multiple solutions. 
So the above model which is a variant of the Penner model with
two cuts, is also conjectured to have multiple solutions.

The potential for a general Penner model is $V(M)=V_0 (M)
-t \ln M$, where $V_0$ is a polynomial. If $V_0 (M) = {1\over 2}
\mu M^2$ the model is the Gaussian Penner model where we can
re-write $\ln M= {1\over 2} \ln M^2$. This has $Z_2$ symmetry
and the potential is a double well with eigenvalues distributed
in disconnected segments. The partition function for the Gaussian
Penner model in terms of its eigenvalues $x_i$ is \cite{penner}

\beq
Z= \int \prod dx_i exp[-E(x)]
\eeq
where $E[x]=N \sum_{i=1}^{N} V(x_i) - 2 \sum_{i\ne j} \ln |x_i-x_j|$
\beq
V(x) = N \left ( {\mu\over 2} x^2 - t \ln |x| \right ).   
\eeq
This potential is similar to that of eq. (\ref{penner}), in that
it has a double-well potential involving $ln |x|$. I 
will show explicitly that this has multiple solutions.  
We consider the situation for $t>0$. 

The recurrence relations for a general Penner model reduce to
\beql
{n\over N} &=& \sqrt{R_n} \langle n-1|V_0^{\prime}(\hat{M})|n\rangle
-t\sqrt{R_n}\langle n-1|\hat{M}^{-1}\|n\rangle \nonumber \\
0 &=& \langle n|V^{\prime}_0 (\hat{M})\|n\rangle 
-t\langle n|\hat{M}^{-1}|n\rangle
\label{pennerrec}
\eeql
Denoting $W_n=\sqrt{R_n}\langle n-1|V_0^{\prime}(\hat{M})\|n \rangle$ and
$Y_n=\langle n|V_0^{\prime}(\hat{M})\|n\rangle$, for the Gaussian Penner
model $W_n=\mu R_n$ and $Y_n=\mu S_n$. We can consider as
in section 4, a period-2 ansatz for $R_n$ and $S_n$ which leads 
to four equations but again only three independent equations:

\beql
C+D &=& 0 \nonumber \\
A+B-CD &=& {{2x+t}\over \mu} \nonumber \\
AB &=& {{x(x+t)}\over \mu^2}.
\label{rec3}
\eeql

Thus there is an infinite class of solutions labelled by one function of 
$x$ in the large $N$ limit. For the `symmetric solution':
$
C = D =0
$ 
and 
$
A = {x\over \mu}, ~~B = {{x+t}\over \mu}
$
while the maximally asymmetric solution:
$
A = B = {1\over \mu}\sqrt{x(x+t)} 
$ and
$
C^2 = {1\over \mu} [(2x+t)-2\sqrt{x(x+t)}].
$
Once again eq. (\ref{rec3}) fixes the same combinations
which appear in the generating function of the Gaussian Penner
model. Thus in the large $N$ limit the eigenvalue density and free energy
are identical for the full infinite class of solutions satisfying 
eq. (\ref{rec3}).

For the symmetric solution 
$\langle n|\hat{M}^{-1}|n \rangle=0$ by $Z_2$ symmetry and $S_n=0$, thus 
eq. (\ref{pennerrec}) yields $R_n={n\over {\mu N}}$ for $n$ even and 
$R_n={n\over \mu N}+{t\over \mu}$ for $n$ odd. This is an exact solution,
hence the exact free energy may be found to be
$
\Gamma=\sum_{k=1}^{N/2-1} k \log[(2k+\mu+1)(2k+\mu-1)],
$
where $t=-1+{\mu\over N}$. Expanding in powers of $\mu$ we get
\beq
\Gamma={1\over 4} \mu^2 \log \mu +{1\over 12} \log \mu \ldots .
\label{sympen}
\eeq
Note that the coefficient of the second term $\log \mu$ is 
$\chi_1={1\over 12}$ which corresponds to the first subleading
correction in the $1\over N$ expansion. For the maximally asymmetric
solution in the double scaled limit $R_n\approx 
{\Gamma(1/2(N-n+\mu+3/2))\over \Gamma(1/2(N-n+\mu+1/2))}$, and the
free energy is $\Gamma=\sum_{k=1}^{N/2-1} 
k \log[(2k+\mu+1/2)(2k+\mu-1/2)]+\ldots$ . On expanding in powers of $\mu$
the free energy is
\beq
\Gamma = {1\over 4} \mu^2 \log \mu - {5\over 48} \log \mu \ldots
\label{asympen}
\eeq
The coefficient of the second term is ${5\over 48}$. Here in
the Gaussian Penner model we see that though the symmetric and 
maximally asymmetric solutions give the same answer in the large 
$N$ limit, the free energies are very different
at higher order. This establishes that in the 
Gaussian Penner model multiple solutions are
present which give the same free energy in the large $N$ limit but
differ at higher orders. As in section 7, certain correlation functions
will be different for these solutions since they depend upon the 
recursion coefficents.

The potential $V(x)$ for the glass model discussed in the high temperature 
region is qualitatively similar to the $V(x)$ for the Gaussian Penner model 
in that both have double wells. This feature is the same for the $M^4$ model 
discussed in earlier sections. Since these later models which have two cuts in
their eigenvalue density have been shown to have multiple solutions,
it is conjectured 
here that the glass model described above in the high temperature phase 
should also have multiple solutions. It would be useful to show this 
explicitly in the future.

More recent work \cite{djb} suggests
that the number of multiple solutions grows
exponentially with $N$. This gives further support for glassy 
behavior in these gapped random matrix models.

\section{Conclusions}
\label{conclusions}

To conclude, ample evidence has been provided for the existence of
multiple solutions in random matrix models with gaps. First a simple
motivation for the presence of multiple solutions is given and then
made more precise using the recurrence coefficients of the orthogonal
polynomials of the system. Further, numerical evidence is 
given for the existence of multiple solutions in this context.
In the large $N$ limit the free energy, generating function and density
are the same. Differences between the multiple solutions are seen in the 
free energy at higher orders in $1\over N$ as well as in correlation 
functions.

Connections with the high temperature phase of structural 
glasses have been made to matrix models \cite{ckpr,p}.
These are a variant of the Penner model with
two-cuts. A simpler model, the Gaussian Penner model, with
disconnected segments is shown to have these unusual multiple solutions
as well. Hence the matrix models with gaps with connections to glass models 
are likely to have this unusual property. The ruggedness of the landscape  
needs to be studied. It would be nice 
to be able to cast these models in the replica framework but 
this remains a difficult task at this point, as the
Hubbard-Stratanovich transformations which are technically needed
for the Gaussian random matrix models ref. \cite{km} are not available 
here for the simple $M^4$ and the Gaussian Penner model or
any other gapped random matrix models. This is a future goal in this
problem.

\section{Acknowledgments}
\label{acknowledgements}

I thank E. Br\'ezin, R. C. Brower, C. Dasgupta,  
S. Jain and C-I Tan for encouragement and collaborations.  
Thanks to S. Jain for critical reading of the manuscript.
I would like to thank the Raman Research Institute, Santa Fe Institute
and the Abdus Salam International Center for Theoretical Physics for 
hospitality and facilities where part of this work was done.


\end{document}